\documentclass[amsmath,amssymb,twocolumn,prl]{revtex4-1}
\usepackage{color}
\usepackage{bm}
\usepackage{graphicx}
\usepackage{epsfig}
\usepackage{amsmath}
\usepackage{amsfonts}
\usepackage{amssymb}
\usepackage{bezier}

\def\beq{\begin{equation}}
\def\eeq{\end{equation}}
\def\be{\begin{equation}}
\def\ee{\end{equation}}
\def\bea{\begin{eqnarray}}
\def\eea{\end{eqnarray}}
\def\half{\mbox{$1\over2$}}

\def\bk{{\bf k}}

\def\bz{{\bf z}}
\def\bn{{\bf n}}
\def\bx{{\bf x}}
\def\by{{\bf y}}

\def\bd{{\bf d}}
\def\bV{{\bf V}}

\def\cH{{ H}}

\def\cT{{\cal T}}

\def\cE{{\cal E}}

\def\bvp{{\bV_\perp(\bk)}}

\def\hbx{{\hat{\bx}}}
\def\hby{{\hat{\by}}}
\def\hbz{{\hat{\bz}}}

\def\hbn{{\hat{\bn}}}
\def\hbd{{\hat{\bd}}}

\def\sd0{\rho^0_s(T)}

\def\1{\mbox{1\hskip-.25em l}}
\def\6{\langle }
\def\9{\rangle }

\def\im{\textrm{Im}}
\def\re{\textrm{Re}}

\def\ve{\varepsilon}

\def\bsg{{\boldsymbol{\sigma}}}

\def\sxyt{C^F}

\def\cbE{\boldsymbol{\cE}}
\def\pk+{\partial_{k+}}

\def\hbnpm{\hbn_\bk}

\def\hbnm{\hbn_\bk}

\def\ccH{\check{\cal{H}}}

\begin{document}
\parindent=0pt

\title{Floquet Topological Insulator in Semiconductor Quantum Wells}
\author{Netanel H. Lindner$^{1,2}$, Gil Refael$^{1,2}$, Victor Galitski$^{3,4}$}
\affiliation{1) Institute of Quantum Information, California Institute of Technology, Pasadena, CA 91125, USA.}
\affiliation{2) Department of Physics, California Institute of Technology, Pasadena, CA 91125, USA.}
\affiliation{3) Condensed Matter Theory Center, Department of Physics,
University of Maryland, College Park, Maryland 20742, USA.}
\affiliation{4) Joint Quantum Institute, Department of Physics,
University of Maryland, College Park, Maryland 20742, USA}
\begin{abstract}
Topological phase transitions between a conventional insulator and a state of matter with topological properties have been proposed and observed in mercury telluride - cadmium telluride quantum wells.
We show that a topological state can be induced in such a device, initially in the trivial phase, by irradiation with microwave frequencies, without closing the gap and crossing the phase transition.  We show that the quasi-energy spectrum exhibits a single pair of helical edge states. The velocity of the edge states can be tuned by adjusting the intensity of the microwave radiation. We discuss the necessary experimental parameters for our proposal. This proposal provides an example and a proof of principle of a new non-equilibrium topological state, Floquet topological insulator, introduced in this paper.
\end{abstract}
\maketitle
\parindent=15pt

Topological phases of matter have captured our imagination over the
past few years, with tantalizing properties such as robust edge modes and exotic non-Abelian excitations \cite{Read, KaneFu}, and potential applications ranging from semiconductor spintronics \cite{Spintronics} to topological quantum computation \cite{RMP}. The discovery of topological insulators in solid-state devices such as HgTe/CdTe quantum wells \cite{Bernevig, Molenkamp}, and in
materials such as $\textrm{Bi}_2\textrm{Te}_3,\,\textrm{Bi}_2\textrm{Sn}_3$ \cite{Hasan, Hasan2, Zhang} brings us closer to employing
the unique properties of topological phases in technological applications.

Despite this success, however, the choice of materials that exhibit
these unique topological properties remains rather scarce. In most
cases we have to rely on serendipity in looking for topological
materials in solid-state structures and our means to engineer
Hamiltonians there are very limited. Therefore, to develop new methods
to achieve and control topological structures at will would be of
great importance.

Our work demonstrates that such new methods
are indeed possible in non-equilibrium,   where external time-dependent
perturbations represent a rich and versatile resource that can be
utilized to achieve topological spectra in systems that are
topologically trivial in equilibrium. In particular, we show that
periodic-in-time perturbations may give rise to new differential
operators with  topological insulator spectra, dubbed Floquet
topological  insulators (FTI), that exhibit chiral edge currents
in non-equilibrium and possess  other hallmark phenomena associated
with topological phases. These ideas, put together with the highly developed technology for controlling low-frequency electromagnetic modes, can enable devices in which fast switching of edge state transport is possible. Moreover, the spectral properties of the edge states, i.e., their
velocity, and the bandgap of the bulk insulator, can be easily controlled. On a less applied perspective, the fast formation of the Floquet topological insulators in response to the external field opens a path to study quench dynamics of topological states in solid-state devices.

The Floquet topological insulators discussed here share many features
discussed in some  previous works: The idea of achieving topological states in a
periodic Hamiltonian was also explored from the perspective of quantum
walks in Ref. \onlinecite{BergRudner}. Also, a similar philosophy led to proposals for the
realization of topological phases in cold-atom systems: a quantum Hall
state using a stroboscopic quadrupole field \cite{SorensenDemler} and a
topological insulator state using a Ramann-scattering induced
spin-orbit coupling \cite{StanescuGalitski}. Also,
Ref. \cite{Oka} proposed to use a circularly-polarized light
field to induce a Hall current in graphene. Another
useful analogy for our work is the formation of zero-resistance state
in Hall bars at low magnetic fields using RF radiation \cite{Mani, Zudov, Aleiner, Assa}.
In our case, it is not the resistance of the bulk that
vanishes, but rather the resistance of the edges, which becomes
finite and universal upon application of the light field. As we were
finalizing our draft, we also became aware of Ref. \onlinecite{Inoue},
which proposed to use a periodic modulation in the form of a
circularly polarized light to change the Chern number in the
Haldane model~\cite{Haldane_M}.

\section{Definition of a Floquet Topological Insulator}

Let us first provide a general construction and definition for a
Floquet topological insulator in a generic lattice model, and then discuss a specific realization: a HgTe/CdTe quantum well. The generic many-body Hamiltonian of interest is
\begin{equation}
\label{Hgen}
\check{\cal H}(t) = \sum_{{\bf k} \in {\rm BZ}} H_{nm} ({\bf k},t) c_{n,{\bf k}}^\dagger c_{m,{\bf k}} + {\rm h.~c.},
\end{equation}
where $c_{n,{\bf k}}^\dagger$ and $c_{m,{\bf k}}$ are fermion creation/annihilation operators,  ${\bf k}$ is the momentum defined in the Brillouin zone, and the Latin indices, $n,m = 1,2,\ldots N$ label some internal degrees of freedom
({\em e.g.}, spin, sublattice, layer indices, etc.). The $N \times N$ ${\bf k}$-dependent matrix $\check{H}({\bf k},t)$ is determined by lattice hoppings and/or external fields, which are periodic in time, $\check{\cal H}(T+t) =\check{\cal H}(t) $.

First, we recall that without the time-dependence, the topological classification reduces to an analysis of the matrix function, $\check{H}({\bf k})$, and is determined by its spectrum.  \cite{class, Kitaev}. An interesting question is whether a topological classification is possible in non-equilibrium, {\em i.e.}, when the
single-particle Hamiltonian, $\check{\cH} ({\bf k},t)$, in Eq.~(\ref{Hgen}) does have an explicit time-dependence and whether there are observable physical phenomena associated with this non-trivial topology.
Consider the single-particle Schr{\"o}dinger equation associated with Eq.~(\ref{Hgen}):
\begin{equation}
\label{SPSE}
\left[ \check{H}({\bf k},t) - i \check{I} \partial_t \right] \Psi_{\bf k} (t) = 0, \,\,\mbox{with } \check{H}({\bf k},t) = \check{H}({\bf k},t+T)
\end{equation}
The Bloch-Floquet theory states that the solutions to Eq.~(\ref{SPSE}) have the form $\Psi_{\bf k} (t) = \check{S}_{\bf k}(t) \Psi_{\bf k} (0)$,
where the unitary evolution is given by the product of a periodic unitary part and a Floquet exponential
\begin{equation}
\label{FT}
\check{S}_{\bf k}(t) = \check{\cal P}_{\bf k} (t) \exp \left[ - i \check{H}_F({\bf k}) t \right],\mbox{ with } \check{\cal P}_{\bf k} (t) = \check{\cal P}_{\bf k} (t+T)
\end{equation}
where $\check{H}_F({\bf k})$ is a self-adjoint {\em time-independent} matrix, associated with the Floquet operator $\left[ \check{H}({\bf k},t) - i \check{I} \partial_t \right]$ acting in the space of periodic functions $\Phi(t) = \Phi(t+T)$, where it leads to a time-independent eigenvalue
problem, $\left[ \check{H}({\bf k},t) - i \check{I} \partial_t \right] \Phi({\bf k}, t) = \ve ({\bf k}) \Phi({\bf k}, t)$. The quasi-energies $\ve ({\bf k})$ are the eigenvalues of the matrix  $\check{H}_F({\bf k})$ in Eq.~(\ref{FT}), and in the cases of interest can be divided into separate bands. The full single-particle wave-function is therefore given by $\Psi(t) = e^{-i \ve t} \Phi(t)$. Note that the quasi-energies are defined modulo the frequency $\omega=2\pi/T$.

The Floquet topological insulator is defined through the topological properties of the time-independent Floquet operator $\check{H}_F({\bf k})$,  in accordance with the existing topological classification of equilibrium Hamiltonians \cite{class,Kitaev}. Most importantly, we show below that the FTI is not only a mathematical concept,
but it has immediate physical consequences in the form of robust edge states that appear in a finite system in the non-equilibrium regime. The density profile of  these hallmark boundary modes was found to be only weakly time-dependent [through the envelope function determined by the matrix $\check{\cal P}_{\bf k} (t) $ in Eq.~\ref{FT}]. Furthermore, we explicitly demonstrate within a specific model that not only topological insulator properties may survive in non-equilibrium,  but they can be induced via a simple non-equilibrium perturbation in an otherwise topologically trivial system.  Using the specific model describing a HgTe/CdTe quantum wells, we show that a carefully chosen non-equilibrium perturbation may be utilized to turn the topological properties on and off. We discuss various methods to realize such a perturbation using experimentally accessible electromagnetic radiation in the microwave-THz regime.

\section{Topological Transition in $\textrm{HgTe}/\textrm{CdTe}$ Heterostructures}

Below we outline a proposal for the realization of a FTI in Zincblende
structures such as the HgTe/CdTe heterostructure which are in the
trivial phase ($d<d_c$ in Ref. \onlinecite{Bernevig}). Consider the Hamiltonian
\beq
\cH(k_x,k_y)= \left(\begin{array}{cc} \check{H}(\bk) &  0 \\ 0 & \check{H}^*(-\bk)
\end{array} \right),
\label{eq: 4by4}
\eeq
where
\beq
\check{H}(\bk)=\epsilon(\bk) \check{I} + \bd(\bk)\cdot \check{\bsg},
\label{eq: 2by2}
\eeq
$\bk=(k_x,k_y)$ is the two dimensional wave vector, and $\check{\bsg}=(\check{\sigma}_x,\check{\sigma}_y,\check{\sigma}_z)$ are the Pauli matrices. The vector, $\bd(\bk)$,
is an effective spin-orbit field. Equations~(\ref{eq: 4by4}) and~(\ref{eq: 2by2}) represent the effective Hamiltonian of HgTe/CdTe heterojunctions \cite{Bernevig}. In these systems, the upper block $\check{H}(\bk)$ is spanned by states with $m_J=(1/2,3/2)$, while the lower block is spanned by states with $m_J=(-1/2,-3/2)$.  The lower block $\check{H}^*(-\bk)$ is the time reversal partner of the upper block, and in the following we shall focus our attention to the upper block of Eq.~(\ref{eq: 4by4}) only.

 The Hamiltonian (\ref{eq: 2by2}) is diagonalized via a ${\bf
   k}$-dependent $SU(2)$ rotation, that sets the z-axis for the
 pseudo-spin along the vector $\bd(\bk)$.
There are two double-degenerate bands  with energies $\epsilon_\pm(\bk) = \epsilon(\bk) \pm \left| {\bf d}({\bf k}) \right|$.
Within each sub-block, the TKNN formula provides the sub-band Chern
number~\cite{TKNN}, which for the Hamiltonian~(\ref{eq: 2by2}) can be
expressed as an integer counting the number of times the vector
$\hbd(\bk)$ wraps around the unit sphere as $\bk$ wraps around the entire FBZ. In integral form, it is given by
\beq
C_\pm=\pm \frac{1}{4\pi}\int d^2\bk \;\hbd(\bk)\cdot \left[ \partial_{k_x}\hbd(\bk)\times
\partial_{k_y}\hbd(\bk)\right],
\label{eq: hall}
\eeq
where $\hat{\bf d}({\bf k}) = {\bf d}({\bf k})/\left| {\bf d}({\bf k}) \right|$ is a unit vector and the $(\pm)$~indices label the two bands.

This elegant mathematical construction yields also important physical consequences, as it is related to the quantized Hall conductance associated with an energy band,
\begin{equation}
\label{xy}
\sigma_{xy} = {e^2 \over h} C.
\end{equation}

Around the $\Gamma$ point of the first Brillouin zone (FBZ) we can expand the vector $\bd(\bk)$ as \cite{Bernevig,Novik}
\beq
\bd(\bk)=\big(A  k_x, A  k_y, M - B\bk^2\big),
\label{eq: bandapprox}
\eeq
where the parameters $A<0,\,B>0$ and $M$ depend on the thickness of the quantum well and on parameters of the materials.
We can easily see that the Chern number implied by $\bd(\bk)$ depends
crucially on the relative sign of $M$ and $B$. Within the
approximation of Eq. (\ref{eq: bandapprox}), far away from the
$\Gamma$ point, $\bd(\bk)$ must point south (in the negative $z$
direction). At the $\Gamma$ point, $\bd(\bk)$ is pointing north for
$M>0$, but south for $M<0$. For the simplified band structure, the
Chern numbers are clearly $C\pm=\pm\left[1+\mbox{sign}(M/B)\right]/2$. For a generic
band structure corresponding to Eq. (\ref{eq: bandapprox}) near the
$\Gamma$-point, the same logic applies, and we can easily see that a
change of sign in $M$ induces a change of the Chern number, $C$, by
$1$.

\begin{figure}
\begin{center}
\vspace{-1.0cm}
\includegraphics[width=11.2cm]{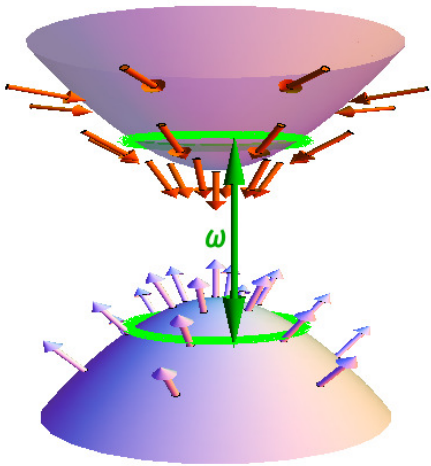}\\
\vspace{-1.0cm}
\caption{Energy dispersion $\epsilon(\bk)$ and pseudospin configuration $-\hbd(\bk)$ for the original bands of $\check{H}(\bk)$ in the non topological phase ($M/B<0$). The non-topological phase is characterized by a spin-texture which does not wrap around the unit sphere. Upon application of a periodic modulation of frequency $\omega$ bigger than the band gap, a resonance appears; the green circles and arrow depict the resonance condition.}
\label{fig: resonance}
\end{center}
\end{figure}

We now show that a similar non-trivial topological structure can be induced in such
quantum wells, starting with the non-topological phase, via periodic modulation of the Hamiltonian,
which allows transitions between
same-momentum states with energy difference of $\hbar\omega$. This creates a circle in the FBZ where transitions between the valence and conduction band are at resonance (see
Fig.~\ref{fig: resonance}). We intend to use the modulation to reshuffle the spectrum such that the resulting valence band
consists of two parts: the original valence band outside the resonance circle
drawn in Fig.~\ref{fig: resonance}, and the original conduction
band inside the resonance circle, near the $\Gamma$ point. From
Fig.~\ref{fig: resonance}, we see that this indeed leads to the desired structure, with the reshuffled pseudospin configuration pointing south near the $\Gamma$ point and north at large k-values (for $M<0$). On the resonance
circle, we expect an avoided crossing separating the reshuffled lower band from the upper band.

\section{Floquet Topological Insulator in a Non-Equilibrium $(Cd,Hg)Te$ Heterostructure}

Let us next consider the Floquet problem in a Zincblende spectrum in
detail. We add a time dependent field to the Hamiltonian~(\ref{eq: 2by2})
\beq
\check{V}(t)=\bV\cdot \check{\bsg}  \cos(\omega t),
\label{eq: periodic}
\eeq
where $\bV$ is a three-dimensional vector, which has to be carefully chosen to
obtain the desired result.
It is convenient to transform the bare Hamiltonian to a ``rotating frame
of reference'' such that the bottom band is shifted by $\hbar
\omega$. This is achieved
using the unitary transformation
$\check{U}(\bk,t)=\check{P}_+(\bk)+\check{P}_-(\bk)e^{i\omega t}$, where
$\check{P}_\pm(\bk)=\half \left[ I\pm\hbd(\bk) \cdot \check{\bsg} \right]$ are projectors on the of upper
and lower bands of $H(\bk)$. This results in the following Hamiltonian:
\beq
\check{H}_I(t)= \check{P}_+(\bk)\epsilon_+(\bk)+ \check{P}_-(\bk)\left[ \epsilon_-(\bk)+\omega \right] +\check{U} \check{V}(t) \check{U}^\dag,
\label{eq: interaction}
\eeq
where $\epsilon_\pm(\bk)$ are the energies corresponding to $\check{P}_\pm(\bk)$. In
the ``rotating'' picture, the two bands cross if $\omega$ is
larger then the gap $M$. The second term in the right-hand-side of Eq.~(\ref{eq: interaction})
is the driving term, which directly couples the bands and has a
time-independent component.

The solution of $H_I$ can also be given in terms of a spinor pointing
along a unit vector,
$\hbn_\bk$, which will play the same role as $\hbd(\bk)$ for the
stationary $\check{H}(\bk)$. $\hbn_\bk$ will encode the topological
properties of the FTI.

\begin{figure}
\begin{center}
\vspace{-4cm}
\includegraphics[width=9.2cm]{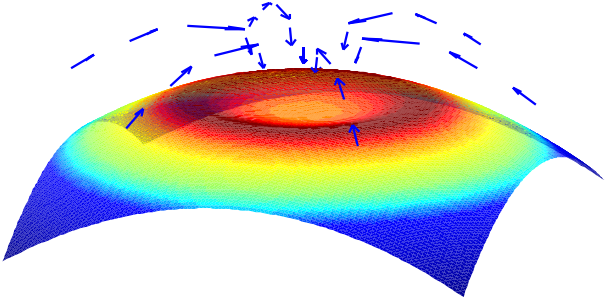}
\vspace{-4.8cm}
\caption{Pseudospin configuration $\hbnm$ (blue arrows) and dispersion of the lower band of $H_I$. Note the dip in the energy surface near $\bk=0$, resulting from the reshuffling of the lower and upper bands of $\check{H}(\bk)$.}
\label{fig: lowerband}
\end{center}
\end{figure}

$H_I$ is solved by the eigenstates $|\psi_I^\pm(\bk)\9$, which for the values of momenta, ${\bf k}$, away from the resonance ring are only weakly
modified compared to the equilibrium, $V=0$, case. We define the vector $\hbnpm=\6 \psi_I^-(\bk)| \check{\bsg} |\psi_I^-(\bk)\9$, which characterizes the pseudospin  configuration in the lower $(-)$ band of $H_I$. The vector $\hbnm$ is plotted in Fig.~\ref{fig: lowerband} for $M/B<0$. Indeed, near the $\Gamma$ point we see that $\hbnm$ points towards the south pole, and for larger values of $\bk$, the band consists
of the original lower band, and therefore $\hbnm$ points towards the northern hemisphere for these $\bk$ values.
These two regimes are separated by the resonance ring, denoted by $\gamma$, for which $\omega=e^+(\bk)-e^-(\bk)$ (the green curve in Fig.~\ref{fig: resonance}).

The topological aspects of the reshuffled lower band depend crucially
on the properties of $\hbnm$ on $\gamma$, which are, in turn,
inherited from the geometric properties of the driving potential,
encoded in $\bV$. These are best illustrated by employing the rotating wave approximation, as we shall proceed to do below. An exact numerical solution will be presented in the next section.

\begin{figure}
\vspace{-0.3cm}
\begin{center}
\includegraphics[width=9.2cm]{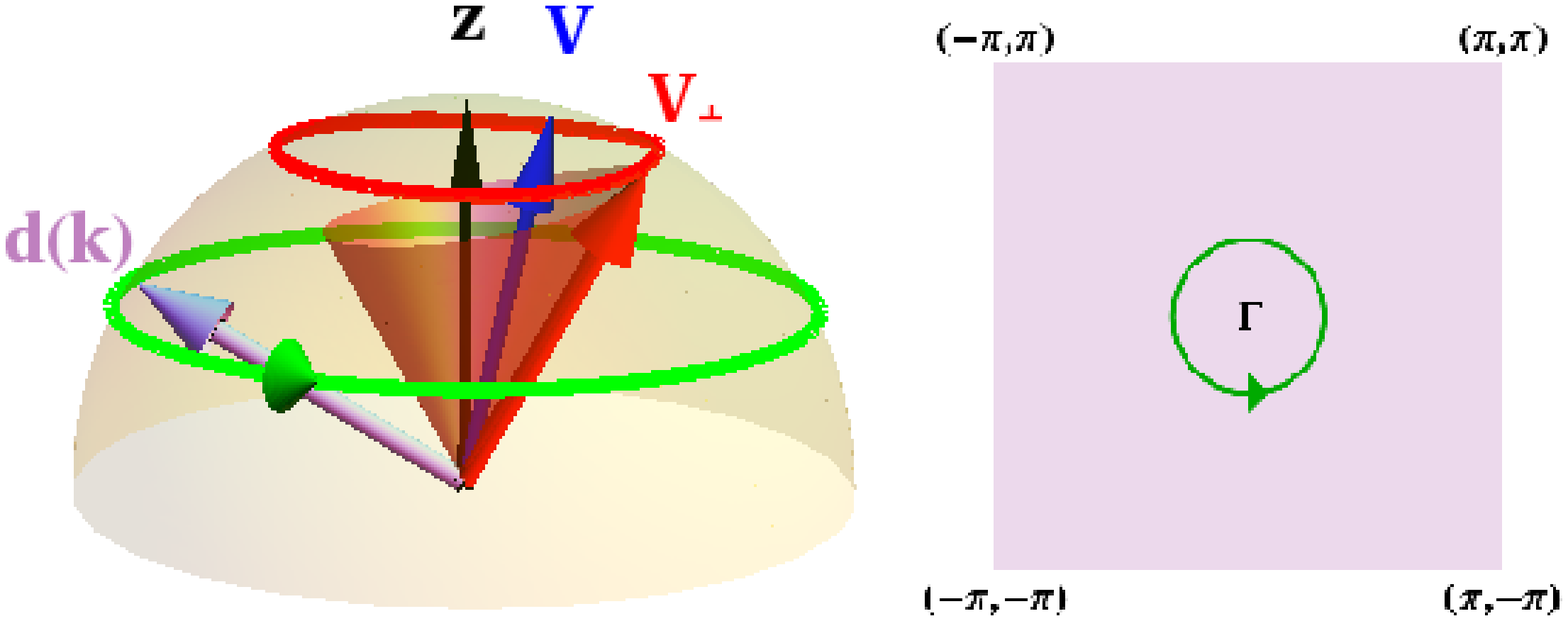}\\
\vspace{-3.0cm}
\caption{The geometrical condition for creating topological quasi-energy bands. The green arrow and circle depicts $\hbd(\bk)$ on the curve $\gamma$ in the FBZ for which the resonant condition holds. The red arrow and cone depicts $\bvp$  on $\gamma$. The blue arrow depicts the driving field vector $\bV$. As long as $\bV$ points within the loop traced by $\hbd(\bk)$, the vector $\bvp$ winds around the north pole, which is indicated by the black arrow.}
\label{fig: geom}
\end{center}
\end{figure}

The driving field $\check{V}(t)$ contains both counter-rotating and co-rotating terms. In the rotating wave approximation it is given by
\beq
\check{V}_{\rm RWA}= \check{P}_+(\bk) \left( \bV\cdot \check{\bsg} \right) \check{P}_-(\bk)+ \check{P}_-(\bk) \left( \bV\cdot \check{\bsg} \right) \check{P}_+(\bk).
\label{eq: vrwa}
\eeq
Next, we decompose the vector $\bV$ as follows
\beq
\bV=\left( \bV\cdot\hbd(\bk) \right)\hbd(\bk)+\bV_\perp(\bk).
\label{eq: vperp}
\eeq
A substitution in Eq.~(\ref{eq: vrwa}) gives
\beq
\check{V}_{\rm RWA}=\bV_\perp \cdot \check{\bsg}.
\label{eq: vrwap}
\eeq
On the curve $\gamma$ we have:
\be
\hbnpm=-\bV_\perp(\bk)/|\bV_\perp(\bk)|.
\ee
We can define a topological invariant $\sxyt$ similar to $C$ in
Eq.~(\ref{eq: hall}), by replacing $\hbd(\bk)$ with $\hbnm$. In order
for $\hbnm$ to have a non-vanishing $\sxyt$, it needs to wrap around
the unit sphere. A necessary condition is that on the curve $\gamma$,
it forms a loop that winds around the north pole. Now, $\bvp$ lies on the plane defined by $\hbd(\bk)$ and $\bV$. For the values of $\bk$ which are on-resonance, $\hbd(\bk)$ traces a closed loop on the unit sphere which encircles the north pole. Therefore, if the driving field vector $\bV$ points to a point on the Bloch sphere which is encircled by this loop, $\bvp$ will also trace (a different) loop on the sphere which encircles the north pole, as illustrated by Fig.~\ref{fig: geom}.

 Under the conditions stated above and with $M<0$, the vector field $\hbnm$ starts from the south pole at $\Gamma$ point and continues smoothly to the northern hemisphere for larger values of $|\bk|$ while winding around the equator. For values of $\bk$ further away from the curve $\gamma$, $\hbnm\approx-\hbd(\bk)$ as the driving field is off resonance there. The contribution of these $\bk$'s to $\sxyt$ is therefore equal to their contribution to $C$. Therefore it is evident that $\sxyt_\pm=C_\pm\pm1$. Note that for $M>0$, $\sxyt_\pm=C_\pm\mp 1$.

A comment is in order regarding the time dependence of $\sxyt$. Since the solutions to the time dependent Schr\"odinger equation are given by transforming back to the Schr\"odinger picture,
\beq
|\psi^\pm(t,\bk)\9=U(t)|\psi^\pm_I(\bk)\9,
\eeq
the pseudospin configuration in the Brillouin zone of these solutions ,
\beq
\hbn_{\bk}(t)= \6 \psi^-(\bk,t) | \check{\bsg}   |\psi^-(\bk,t) \9,
\label{eq: nk}
\eeq
will also depend on time. As long as $H_I$ is non-degenerate in the FBZ, which implies that $\hbn_{\bk}(t)$ is well defined, $\sxyt$ will be quantized to an integer. As for $C$, also $\sxyt$ is a topological invariant which is robust to smooth changes in $\hbnm(t)$ which are not singular at any point in the FBZ. Therefore, $\sxyt$ \textit{does not} depend on time, although the pseduospin configurations $\hbnm(t)$ do, and we can calculate $\sxyt$ using $\hbnm$.

\section{Non-Equilibrium Edge States}
\label{sec: edge}

One of the most striking results of the above considerations is
the existence of helical edge states once the time dependent field is turned on.
Below we demonstrate the formation of edge states in a tight binding model which contains the essential features of Eq.~(\ref{eq: 2by2}).
The Fourier transform of the spin-orbit coupling vector, ${\bf d} ({\bf k})$ in the corresponding lattice model is given by, c.f., Eq.~(\ref{eq: bandapprox}),
\begin{eqnarray}
{\bf d} ({\bf k}) = \left( A \sin k_x,\, A \sin k_y,\, M-4B+2B [\cos k_x +\cos k_y] \right).
\nonumber \\
\phantom{1}
\label{eq: band}
\end{eqnarray}
We consider the above model with the time dependent field of the form $V_0\check{\sigma}_z \cos(\omega t)$ in the strip geometry, with periodic boundary condition in the $x$ direction, and vanishing boundary conditions at $y=0,L$.

\begin{figure}
\begin{center}
\vspace{-3.5cm}
\includegraphics[width=12cm]{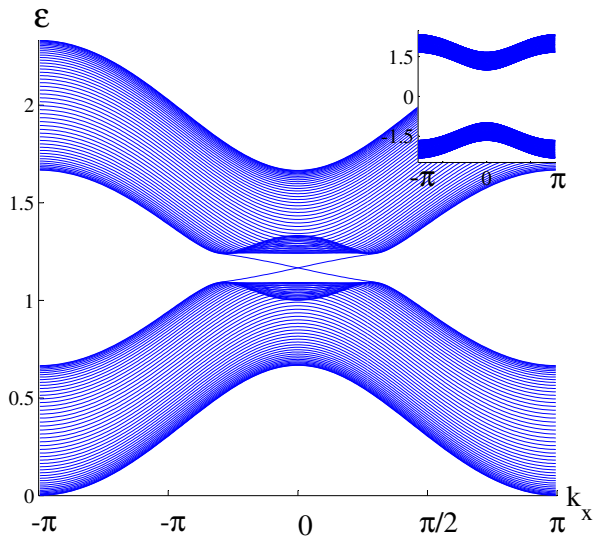}
\vspace{-4.5cm}
\caption{Quasi-energy spectrum of the Floquet equation (\ref{FT}) of the Hamiltonian ~(\ref{eq: band}), in the strip geometry: periodic boundary conditions in the $x$ direction, and vanishing ones in the $y$ direction. The driving field was taken to be in the $\hbz$ direction. The horizontal axis labels the momentum $k_x$. The vertical axis labels the quasi-energies in units of $|M|$. Two linearly dispersing chiral edge modes traverse the gap in the quasi-energy spectrum. The parameters used are $\omega=2.3|M|$, $|\bV|=A=|B|=0.2|M|$. The inset shows the dispersion of the original Hamiltonian ~(\ref{eq: band}), for the same parameters.}
\label{fig: edgestates}
\end{center}
\end{figure}

We solve the Floquet equation numerically by moving to frequency space and truncating number of harmonics.
The wave vector $k_x$ is therefore a good quantum number, and the
solutions $\Phi(t)$ are characterized by $\ve$ and $k_x$.
The quasi-energies for this geometry is displayed in Fig.~\ref{fig: edgestates}.
The quasi-energies of the bottom and top band represent
modes which are extended spatially, while for each value of $k_x$
there are two modes which are localized in the $y$ direction.

As is evident from Fig.~\ref{fig: edgestates}, the \textit{quasi-energies}
of these modes disperse linearly, $\ve(k_x)\propto k_x$, hence they are propagating
with a fixed velocity. Consider a wave packet which is initially
described by $f_0(k_x)$. From equation~(\ref{FT}) we
see that it will evolve into $\psi(t)=\int dk_x
e^{i\ve(k_x)t}f_0(k_x)\Phi^{e}_{k_x}(y,t)$, where $\Phi^{e}_{k_x}$
denotes the quasi energy \textit{edge} states with momentum
$k_x$. Clearly, this will give a velocity of $\6\dot{x}\9=\int dk_x
|f(k_x)|^2\frac{\partial \epsilon}{\partial k_x}$.

\begin{figure}
\begin{center}
\vspace{-3.2cm}
\epsfxsize=.50\textwidth \centerline{\epsffile{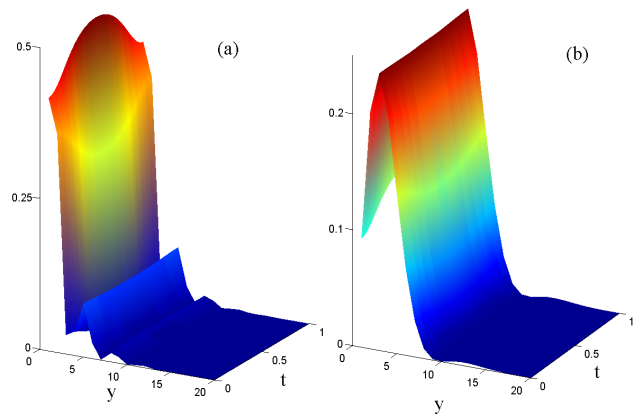}}
\vspace{-3cm}
\caption{Density of edge mode as function of time, $|\phi(y,t)|^2$, (a) for $k_x=0$, and (b) for $k_x=0.84$, where the edge modes meet the bulk states. The horizontal axis display the distance from the edge, $y$, in units of the lattice constant, and the time in units of $2\pi/\omega$. Only the density for the 20 lattice sites closest to the edge are shown for clarity.}
\vspace{-1cm}
\label{fig: density vs time}
\end{center}
\end{figure}

In general, the solutions $\Phi_{\ve,k_x}(t)$ are time-dependent. An
important finding is that the density edge modes are only very weakly
dependent on time. This can be seen in Fig.~\ref{fig: density vs time}, in which we plot the time dependence
of the density profile of these modes.

\section{Experimental Realization of the FTI}

To experimentally realize the proposed state, we need identify a proper
time-dependent interaction in the HgTe/CdTe wells. Below we consider
several options, of which the most promising one uses a periodic
electric field, and the strong linear Stark effect that arises due to
the unique band structure.

{\em Magnetic field realization --} Perhaps the simplest realization of a time dependent perturbation of the form~(\ref{eq: periodic}) is by a microwave-THz oscillating magnetic field, polarized in the $\bz$ direction.
The effect of Zeeman energies in thin Hg/CdTe
quantum wells can be evaluated by recalling that the effective model~(\ref{eq: 4by4})
includes states with $m_J=\pm(1/2,3/2)$ in the upper and lower block respectively. This would result in an effective Zeeman splitting between the two states in each block \cite{Novik}.
The value for the g-factor for HgTe semiconductor quantum wells was
measured to be $g\approx20$ \cite{gfactor}. Therefore, a gap in the quasienergy spectrum on the
order of $0.1 \textrm{K}$ can be achieved using magnetic fields of $10
\textrm{mT}$. Bigger gaps may be achieved by using Se instead of Te, as its g-factor
is roughly twice as large \cite{gfactor2}.

As can be seen by inspecting
Eq.~(\ref{eq: vperp}), the Chern numbers $\sxyt$ for each block in this realization
depend only on the winding of the vector $\hbd(\bk)$. Therefore, the two
blocks will exhibit \textit{opposite} $\sxyt$, resulting in two \textit{counter}-propagating helical edge
modes. As we explain in the next section, the counter-propagating edge modes cannot couple to open a gap in the quasi-energy spectrum, even though a magnetic field is odd under time reversal.

{\em Stress Modulation --} Stress
modulation of the quantum wells using piezo-electric materials, would lead to modulation of the parameter $M$ in~(\ref{eq:  2by2}), and leads to two counter propagating edge states.

{\em Electric field realization --} An in-plane electric field is
perhaps the most promising route to the FTI, can produce large
gaps in the quasi-energy spectrum (compared the
Zeeman case), and leads to robust co-propagating edge modes. The electric field is given by
\beq
\check{E}=\re(\cbE \cdot \exp {i \omega t}) i\nabla_\bk.
\label{eq: E}
\eeq
Inserting this into Eq.~(\ref{eq: vrwa}), we get
\beq
\bvp=\hbd(\bk) \times (\re\cbE\cdot \nabla_{\bk})\hbd(\bk)-(\im\cbE \cdot \nabla_{\bk})\hbd(\bk).
\label{eq: vrwa electric}
\eeq
As before, the vector field $\bvp$ is orthogonal to $\hbd(\bk)$, and again, we would like it to wind around the north pole. Now if we take $\cbE=\cE(-i\hbx+\hby)$ we get, expanding Eq.(\ref{eq: vrwa electric}) to second order in $k_x,k_y$,
\beq
\bvp=\frac{A (A^2 - 4 B M)\cE}{M^3}\Big[\half(k_x^2-k_y^2)\hbx
 + k_x k_y \hby\Big].
\eeq
Evidently, the vector field $\bvp$ winds twice around the
equator. Therefore, for the above choice of $\cbE$, the Chern
numbers will be $\sxyt_\pm=\pm2$. We note that for the lower block will
have $\sxyt=0$. Therefore, each edge of the system will have two
\textit{co}-propagating chiral modes, with spin in the sector of
$m_J=1/2,3/2$. Naturally, a choice of $\cbE=\cE(i\hbx+\hby)$
will give $\sxyt_\pm=\mp2$ for the lower block and $\sxyt=0$ for the upper
block. The evolution with the oscillating electric field is
topologically distinct from the trivial one, as the two co-propagating edge modes cannot become gapped. For
HgTe/CdTe quantum wells with thickness of $58{\AA}$\cite{Bernevig}, we
have $|\bV(\bk)/\cE|\approx 1 \textrm{mm}$ , which leads to a gap in
the quasi-energy spectrum on the order of $10 K$ already for modest
electric fields on the order of $1\frac{\textrm{V}}{\textrm{m}}$ which
are experimentally accessible with powers $<1\textrm{mW}$. We note that decreasing
the well thickness increases \cite{Rothe} the value of $|A/M|$, which
can help achieve even larger gaps in the quasi energy spectrum.

\section{Discussion}
In summary, we showed that the quasi-energy spectrum of an
otherwise ordinary band insulator irradiated by electromagnetic fields
can exhibit non-trivial topological invariants and chiral edge
modes. A realization of these ideas in Zincblende systems, such as
HgTe/CdTe semiconducting quantum wells, can lead to Floquet Topological Insulators that support either co- or
counter-propagating helical edge modes. The Floquet operators of these realizations belong, respectively, to symmetry classes analogous to classes A (no symmetry) and AII (time reversal symmetry with $\cT^2=-1$) in \cite{class}.

The symmetry class of the Floquet Topological Insulator indeed requires careful consideration when two counter-propagating edge states are present, as in the oscillating magnetic-field realization suggested in the previous section. In time independent systems, topological phases exhibiting counter-propagating edges are only distinct from trivial phases under the restriction $\cT H \cT^{-1}=H$, where $\cT$ is the anti-unitary time reversal operator satisfying $\cT^2=-1$. In the time-periodic case, the Hamiltonian at any given time may not possess any symmetry under time reversal. Nevertheless, when the condition
\beq
\cT \ccH (t) \cT^{-1}=\ccH(-t+\tau)
\label{eq: condition}
\eeq
holds (for some $\tau$), the Floquet matrix of Eq.~(\ref{FT}) satisfies $\tilde{\cT} \check{H}_F(\bk) \tilde{\cT}^{-1}=\check{H}_F(-\bk)$, where $\tilde{\cT}$ is an anti-unitary operator which is related to $\cT$ by $\tilde{\cT}=V^{\dag}\cT V$, with $V=\check{S}_\bk\left(-(T+\tau)/2\right)$, c.f. Eq.~(\ref{FT}). Clearly, $\tilde{\cT}^2=-1$.  Therefore, under this condition, the quasi-energy spectrum consists of analogues to Kramer's doublets, which cannot be coupled by the Floquet matrix. The counter propagating edge-modes are such a Kramer's pair, which, therefore, cannot couple and open a gap (in the quasi-energy spectrum) under any perturbations satisfying Eq.~(\ref{eq: condition}). We note that Eq.~(\ref{eq: condition}) holds for any Hamiltonian of the form $\ccH(t)=\ccH_0+\check{V}\cos(\omega t + \phi)$, with time reversal invariant $\ccH_0$, and $\check{V}$ having unique parity under time reversal, \textit{i.e.}, $\cT\check{V}\cT^{-1}=\pm\check{V}$. An oscillating magnetic field, being odd under time reversal, therefore obeys Eq.~(\ref{eq: condition}) and leads to two counter propagating edge modes.

Another important question that we did not touch upon is the
 non-equilibrium onset and steady states~\cite{AR_VG} of the driven systems.
We emphasize that in the presence of the time dependent fields, response
functions including Hall conductivity will be determined not only by the
spectrum of the Floquet operator, but also by the distribution of
electrons on this spectrum. These in turn depend on the specific
relaxation mechanisms present in the system, such as electron-phonon
mechanisms \cite{eliashberg,eliashberg2} and electron-electron interaction \cite{glazman}. An interesting observation here is that the
average energies of the edge modes lie in the original bandgap.

One way to minimize the unwanted non-equilibrium heating effects would be to use
an adiabatic  build-up of the Floquet topological insulator state, e.g. with the
 frequency of the modulation gradually increasing from zero. This should result,
 at least initially, in a filled lower Floquet band.  Nevertheless, the scattering events will
 lead to a relaxation from the conduction band into the valence band of the
 original system and will always play  role by producing mobile bulk quasi-particles. Perhaps this relaxation
 could be suppressed by restricting the corresponding optical modes in
 the environment. An analysis of the non-equilibrium states of the
 system will be the subject of future work.

We thank Joseph Avron, Assa Auerbach, Erez Berg, Andrei Bernevig, James Eisenstein, Lukasz Fidkowski, Victor Gurarie, Israel Klich, and Anatoli Polkovnikov for illuminating conversations. This research was supported by
DARPA (GR, VG), NSF grants PHY-0456720 and PHY-0803371 (GR, NL). NL acknowledges the financial support of the
Rothschild Foundation and the Gordon and Betty Moore Foundation.

\end{document}